\documentclass[preprint,showpacs,amsmath,amssymb,nofootinbib,aps]{revtex4}

\renewcommand{\thefootnote}{\alph{footnote}}
\renewcommand{\thesection}{\Roman{section}}
\renewcommand{\theequation}{\thesection.\arabic{equation}}

\newcommand{\pdo}[1]{\ensuremath{\frac{\partial }
        {\partial #1 }}}

\newcommand{\slashletter}[1]{\ensuremath{\kern+0.1em /\kern-0.65em #1}}

\begin{document}

\renewcommand{\thefootnote}{\alph{footnote}}

\title{Mass zeros in the one-loop effective actions of QED in 1+1 and 3+1
dimensions}
\author{M. P. Fry}
\email{mpfry@maths.tcd.ie}
\affiliation{School of Mathematics, University of Dublin, Dublin 2, Ireland}

\begin{abstract}
It is known that the one-loop effective action of $\mbox{QED}_2$ is a
quadratic in the field strength when the fermion mass is zero: all
potential higher order contributions beyond second order vanish.  For
nonzero fermion mass it is shown that this behavior persists for a
general class of fields for at least one value of the fermion mass
when the external field's flux $\Phi$ satisfies $0<|e\Phi|<2\pi$.  For
$\mbox{QED}_4$ the mass-shell renormalized one-loop effective action
vanishes for at least one value of the fermion mass for a class of
smooth, square integrable background gauge fields provided a plausible
zero-mass limit exists.\\
\end{abstract}


\maketitle

\section{INTRODUCTION}
\setcounter{equation}{0}
\renewcommand{\thesection}{\arabic{section}}
\renewcommand{\theequation}{\thesection.\arabic{equation}}

In all gauge field theories coupled to fermions the fermionic
determinant is fundamental.  These determinants, denoted by
$\text{det}$, produce an effective functional measure for the gauge
fields when the fermionic fields are integrated.  The continuing lack
of nonperturbative information on these determinants is reflected in
the necessity to make loop expansions or the more extreme quenched
approximation in which the determinant is ignored.  Nonperturbative
approaches such as Monte Carlo evaluations with a discrete lattice
regulator result in algorithms that currently dominate this area.
Most analytic nonperturbative results obtained so far deal with the
dependence of the determinants on the coupling constant.  Little
attention has been given to their dependence on the fermion mass.
Here we will confine our attention to quantum electrodynamics in two
and four dimensions in the belief that any progress made might suggest
how to proceed in other cases.  Furthermore, nonperturbative
$\text{QED}$ is of interest in its own right.

It might be objected that $\text{QED}_2$ is of no physical interest,
certainly not the mass dependence of its fermionic determinant.  This
is not true.  Firstly, when the Wick rotation to Euclidean space is
made, $\text{det}_{\text{QED}_2}$ is calculated from
the eigenfunctions of the two-dimensional Pauli operator
$(P - A)^2 - \sigma_3B$ in a magnetic field $B$ normal to a plane.  In
what follows the coupling constant $e$ is assumed to be absorbed by
the potential $A_\mu$.  Then $\text{det}_{\text{QED}_2}$ fully
determines $\text{det}_{\text{QED}_3}$~\cite{i} and
$\text{det}_{\text{QED}_4}$~\cite{ii} for the same magnetic field,
namely
\begin{equation}
\frac{\partial}{\partial m^2} \ln \text{det}_{{\text{\scriptsize{QED}}_4}} = - \frac{L^2}{2\pi} \ln
\text{det}_{{\text{\scriptsize{QED}}_2}} - \frac{L^2 \|B\|^2}{24\pi^2m^2}, \label{eqn:ione}
\end{equation}
\begin{equation}
\ln \text{det}_{{\text{\scriptsize{QED}}_3}} = \frac{L}{2\pi} \int^\infty_{m^2}
\frac{dM^2}{\sqrt{M^2 - m^2}}\ \ln \text{det}_{{\text{\scriptsize{QED}}_2}} (M^2), \label{eqn:itwo}
\end{equation}
where $L$ is the edge length of a space-time box and $\|B\|^2 =
\int d^2x\,B^2(x)$.  Equation (\ref{eqn:ione}) assumes mass-shell change
renormalization while (\ref{eqn:itwo}) assumes a $2\times 2$ representation of
the Dirac $\gamma$-matrices.  By continuing back to Minkowski space
these equations give the effective action $iS = \ln \text{det}$ for a
two-variable static magnetic field in 2+1 and 3+1 dimensions.

Secondly, suppose $\text{det}_{{\text{\scriptsize{QED}}_2}}$ is
calculated for the single-variable magnetic field $B =
B_0f(x/\lambda)$.  The duality transformation $B
\to e^{-i\pi/2}E$, where $E = E_0f(t/\tau)$ and $\tau
= i \lambda$, gives the pair nonproduction probability $e^{-2ImS}$ with
$ImS^{3+1}$ and $ImS^{2+1}$ obtained from (\ref{eqn:ione}) and
(\ref{eqn:itwo}).  Duality in this restricted sense has been
demonstrated recently by Dunne and Hall~\cite{iii,iv}.  Conditions
for the validity of the more general duality transformation $B(x,y)
\to e^{-i\pi/2} E(x,t)$ are unknown.

There are no exact calculations of $S$ in any dimension for
two-variable fields $B(x,y)$ or $E(x,t)$, or even finite-flux magnetic
fields, except for the two-dimensional case of a magnetic field
confined to the wall of a cylinder~\cite{v}.  Actions for slowly
varying fields can be calculated in a derivative
expansion~\cite{vi,vii,viii,ix,x}.  For more general fields
semiclassical estimates of $S$ are effective provided the analysis can
be carried through~\cite{iii,xi}.  So far this has limited the
background field to a dependence on a single space or time variable,
effectively special cases of $\text{QED}_2$.

The Euclidean $\text{QED}_2$ determinant can be expressed
as~\cite{ii,xii,xiii,xiv}

\vspace{2mm}
\begin{equation}
\ln \text{det}_{{\text{\scriptsize{QED}}_2}} = -\frac{1}{2\pi} \int \frac{d^2k}{(2\pi)^2}\
|\hat{B}(k)|^2 \int^1_0 dz\ \frac{z(1-z)}{k^2z(1-z) + m^2} + \ln \text{det}_3,
\label{eqn:ithree}
\end{equation}

\vspace{2mm}
\noindent
where $\hat{B}$ is the Fourier transform of B and $\ln \text{det}_3$
may for the present be viewed as the sum of all one-loop fermion
graphs, beginning in fourth order.  It is gauge invariant, depending
only on $B$.  It is known that $\ln \text{det}_3(m^2\!\!=\!\!0) = 0$.
This was first shown by Schwinger~\cite{xv}.  Seiler~\cite{xii} later
gave a compact proof of Schwinger's result and stated the precise
condition for it to be true, namely
$A_\mu \in \bigcap_{n>2} L^n(\text{I}\!\text{R}^2)$. Furthermore,

\vspace{2mm}
\begin{equation}
\lim_{m^2 = 0} \ln \text{det}_3(m^2) = 0,
\label{eqn:ifour}
\end{equation}

\vspace{2mm}
\noindent
provided the magnetic field's flux $\Phi = 0$.  This result requires
several nontrivial estimates from analysis and will be published
elsewhere.  But it is plausible:  if $\hat{B}(0) = 0$ or equivalently,
$\Phi = 0$, then the infrared properties of $\ln \text{det}_3$ are improved,
allowing continuity at $m^2 = 0$.  In Sec. \ref{sec:two} we will show
that for potentials $A_\mu \in \bigcap_{n>2} L^n(\text{I}\!\text{R}^2)$
and finite-range magnetic fields with $B \in L^n(\text{I}\!\text{R}^2)$,
 $n = 2, 4$ and $\int d^2x\,B^2(\partial_\mu B)^2 < \infty$ there is
at least one value of $m^2>0$ for which $\ln \text{det}_3 = 0$,
provided $0<|\Phi|<2\pi$. Therefore, our result is this:  when
$0<|\Phi|<2\pi$ the zero in $m^2$ of $\ln \text{det}_3$ moves up from
$m^2 = 0$ when $\Phi=0$ to some finite value(s) $m^2>0$.  For
$|\Phi|\geq 2\pi$ our analysis is unable to say anything about the
zeros in $m^2$ of $\ln \text{det}_3$.  Apparently their presence or
absence is tied in with the formation of square-integrable zero modes
of the two-dimensional Pauli operator when $|\Phi| > 2\pi$~\cite{xvi}.

The presence of zeros in $m^2$ in $\ln \text{det}_3$ together with the
result~\cite{xvii} that $\ln \text{det}_3$ is bounded above and below by
terms quadratic in $B$ suggest that $\ln \text{det}_3$ is small in
the sense
that it is comparable to the second-order term in (\ref{eqn:ithree}).
These bounds are obtained from Eq.(9) in~\cite{xvii} and the
definition (\ref{eqn:ithree}) above.  The lower bound on $\ln \text{det}_3$
from~\cite{xvii} has been established for fields $B \geq 0$ or
$\leq 0$ over all space, a technicality that a better estimate might
overcome.

In Sec. \ref{sec:three} we establish the conditions for the large
$m^2$ expansion of $\ln \text{det}_3$ to be an asymptotic series, a result
required in Sec. \ref{sec:two} and useful in Sec. \ref{sec:four}.

For Euclidean $\text{QED}_4$ we will present evidence in
Sec. \ref{sec:four} that $\ln \text{det}_{{\text{\scriptsize{QED}}_4}}$
vanishes for at least one value of $m^2$ for a class of smooth,
square-integrable background gauge fields $F_{\mu\nu}$ provided
$A_\mu \in \bigcap_{n>4} L^n(\mathbb{R}^4)$. Our result
is tentative as it requires the proof of the limit in
(\ref{eqn:iveight}) below.  We believe (\ref{eqn:iveight}) can be
proved, thereby validating new nonperturbative information on
$\text{QED}_4$.

\renewcommand{\thesection}{\Roman{section}}
\renewcommand{\theequation}{\thesection.\arabic{equation}}
\section{TWO-DIMENSIONAL QED} \label{sec:two}
\setcounter{equation}{0}
\renewcommand{\thesection}{\arabic{section}}
\renewcommand{\theequation}{\thesection.\arabic{equation}}

The fermionic determinant in Euclidean QED is defined here by
Schwinger's~\cite{xviii} heat kernel representation

\vspace{2mm}
\begin{equation}
\ln \text{det} = \frac{1}{2} \int_0^\infty \frac{dt}{t} \left\{ \text{Tr}
\left(\!\ e^{-P^2t} - \exp[-(D^2 + \frac{1}{2} \sigma F) t\!\ ]\!\ \right)
+ \frac{\|F\|^2}{24\pi^2} \right\} e^{-tm^2}. \label{eqn:iione}
\end{equation}

\vspace{2mm}
\noindent
Here $D^2 = (P - A)^2$, $\sigma^{\mu \nu} = [\gamma^\mu,
\gamma^\nu]/2i$, $\gamma^{\mu\,\dagger} = -\gamma^\mu$, and $\|F\|^2 =
\int d^4x F_{\mu \nu}^2$.  The last term in (\ref{eqn:iione}) is the
second-order mass-shell charge renormalization subtraction required for
the small $t$ limit of the integral to converge.  In two and three
dimensions this term should be omitted.  If $-\ln \text{det}$ is
combined with the Maxwell action to form an effective measure for
$A_\mu$ then $A_\mu$ has to be concentrated on $\mathcal{S'}$, the
Schwartz space of tempered distributions.  Such potentials have to be
temporarily smoothed until after the integration over $A_\mu$ if sense
is to be made of the right-hand side of (\ref{eqn:iione}).  This
procedure has been discussed elsewhere~\cite{i,v}.  Here we will
simply assume that $A_\mu$ and $F_{\mu\nu}$ are sufficiently smooth
and fall off rapidly enough for our analysis to go through.  More
specific statements will be made below.

Specializing to $\text{QED}_2$ and expanding the right-hand side
of (\ref{eqn:iione}) to second order gives the standard perturbative
result in the first term in (\ref{eqn:ithree}).  The remainder,
$\ln \text{det}_3$, is given by (\ref{eqn:iiitwo}) below.  In
coordinate space (\ref{eqn:ithree}) is

\vspace{2mm}
\begin{equation}
\ln \text{det}_{\text{\scriptsize{QED}}_2} = \int d^2x\,d^2y\,B(x)\,\Pi (x-y) B(y) + 
\ln \text{det}_3, \label{eqn:iitwo}
\end{equation}
where
\begin{eqnarray}
\Pi(x) &=& -\frac{1}{2\pi}\int \frac{d^2k}{(2\pi)^2} e^{ikx} \int_0^1 dz
\frac{z(1-z)}{k^2 z(1-z) + m^2} \nonumber \\
&=& -\frac{1}{4\pi^2}\int_0^1 dz\, K_0 \left(\frac{|mx|}{\sqrt{z(1-z)}}\right).
\end{eqnarray}

\vspace{2mm}
\noindent
Assuming that B has finite range, the $m^2 \to 0$ limit can be
interchanged with the $x$ and $y$ integrals in (\ref{eqn:iitwo}),
giving

\vspace{2mm}
\begin{equation}
\int d^2x\,d^2y\,B(x)\,\Pi(x-y) B(y)\ 
\lower7pt \hbox{$\scriptscriptstyle m^2 \downarrow 0$}
\mkern-25mu =\ 
\frac{\Phi^2}{8\pi^2}
\ln m^2 + O(1). \label{eqn:iifour}
\end{equation}

\vspace{2mm}
\noindent
We have shown  that if $B$ is square integrable and has finite range
then~\cite{xix}

\vspace{2mm}
\begin{equation}
\ln \text{det}_{{\text{\scriptsize{QED}}_2}}\ 
\lower7pt \hbox{$\scriptscriptstyle m^2 \downarrow 0$}
\mkern-25mu = \ 
\frac{|\Phi|}{4\pi} \ln m^2 + R(m^2), \label{eqn:iifive}
\end{equation}

\vspace{2mm}
\noindent
where 
$ \lower7pt \hbox{$\scriptscriptstyle m^2 \downarrow 0$}
\mkern-30mu \lim\ [R(m^2)/\ln m^2] = 0$.  That 
$\ln \text{det}_{{\text{\scriptsize{QED}}_2}}$ is negative is a
reflection of the paramagnetic property of charged fermions whereby
the eigenvalues of the Pauli operator are on average lower relative to
those of the free Hamiltonian $P^2$ in the definition
(\ref{eqn:iione})~\cite{xx,xxii}.  The mass singularity in 
$\ln \text{det}_{{\text{\scriptsize{QED}}_2}}$ at $m^2=0$ is due to
the formation of square-integrable zero modes and zero-energy unbound
resonances in the continuum part of the Pauli operator's spectrum.
The difference between (\ref{eqn:iifour}) and (\ref{eqn:iifive}) makes
the nonperturbative nature of the result (\ref{eqn:iifive}) evident.
Equations (\ref{eqn:iitwo}), (\ref{eqn:iifour}) and
(\ref{eqn:iifive}) give

\vspace{2mm}
\begin{equation}
\ln \text{det}_3\ 
\lower7pt \hbox{$\scriptscriptstyle m^2 \downarrow 0$}
\mkern-25mu = \ 
\frac{|\Phi|}{4\pi} \left( 1 - \frac{|\Phi|}{2\pi} \right) \ln m^2 + R(m^2),
\label{eqn:iisix}
\end{equation}

\vspace{2mm}
\noindent
from which one infers that $\ln \text{det}_3<0$ if $0<|\Phi|<2\pi$
and $m^2$ is sufficiently small.

The second piece of nonperturbative information required is 

\vspace{2mm}
\begin{equation}
\ln \text{det}_3
\lower7pt \hbox{$\scriptscriptstyle m^2 \to \infty$} 
\mkern-25mu =\ 
\frac{1}{90\pi m^6} \int d^2x\, B^4 + {\mathcal{R}} (m^2),
\label{eqn:iiseven}
\end{equation}

\vspace{2mm}
\noindent
where
\begin{equation}
\lim_{m^2 \to \infty} m^6 {\mathcal{R}}  (m^2) = 0.
\label{eqn:iieight}
\end{equation}

\vspace{2mm}
This will be shown in Sec. \ref{sec:three}.  It shows that for
sufficiently large $m^2$\ \ $\ln \text{det}_3$ becomes positive before
approaching zero.  This establishes our claim that $\ln \text{det}_3$
has at least one zero for $m^2 > 0$ when $0 < |\Phi| < 2\pi$.

\renewcommand{\thesection}{\Roman{section}}
\renewcommand{\theequation}{\thesection.\arabic{equation}}
\section{LARGE MASS BEHAVIOR OF \protect\(\text{det}_3 \protect\)} \label{sec:three}
\setcounter{equation}{0}
\renewcommand{\thesection}{\arabic{section}}
\renewcommand{\theequation}{\thesection.\arabic{equation}}

Here we will demonstrate (\ref{eqn:iiseven}) and (\ref{eqn:iieight}).
Integration over the fermions produces the formal result $\text{det} (
\! \not\!\! P - \! \not\!\! A + m)/\text{det}(\! \not\!\! P + m).$
Another formal operation reduces this to $\text{det}(1 - S \!
\not\!\!\! A)$,
where $S = (\! \not\!\! P + m)^{-1}$.  Because neither $S\! \not\!\!\! A$ nor
$(S\! \not\!\!\! A)^2$ are trace class while $(S\! \not\!\!\! A)^3$ is in
$\text{QED}_2$ (see below) the identity $\ln \text{det} (1+A) =
\text{Tr} \ln(1+A)$ for trace class operators has to be modified to

\vspace{2mm}
\begin{equation}
\ln \text{det}_3 (1 - S \! \not\!\! A) = 
\text{Tr}\,[\, \ln(1-S\! \not\!\! A) + S \! \not\!\! A + \frac{1}{2}(S\! \not\!\! A)^2 ].
\label{eqn:iiione}
\end{equation}

\vspace{2mm}
\noindent
The right-hand side of (\ref{eqn:iiione}) is the standard definition
of a regularized determinant~\cite{xiii,xxiii,xxiv,xxv,xxvi}.  
Since $\text{Tr}\,(S \! \not\!\!\! A)^3 = 0$ by Furry's theorem, the first
nonvanishing term in (\ref{eqn:iiione}) begins in fourth order.  This
leaves the second-order term in $\ln \text{det}_{{\text{\scriptsize{QED}}_2}}$
to be defined by expanding definition (\ref{eqn:iione}) to second
order, giving (\ref{eqn:ithree}).  Subtracting the second-order term
from the heat kernel representation (\ref{eqn:iione}) of 
$\ln \text{det}_{{\text{\scriptsize{QED}}_2}}$ gives a definition of 
$\ln \text{det}_3$ equivalent to (\ref{eqn:iiione})~\cite{xiii}:

\vspace{2mm}
\begin{eqnarray}
\ln \text{det}_3 &=& \frac{1}{2}
\int_0^\infty \frac{dt}{t} \left\{ \text{Tr} \left( e^{-P^2t} - \exp[-(D^2 +
\frac{1}{2} \sigma F) t] \right) \right.
\nonumber \\
&\ &+ \left. \frac{t}{2\pi}\int^1_0 dz\, z\, (1-z)\int\frac{d^2k}{(2\pi)^4} 
\,|\hat{F}_{\mu\nu} (k) |^2\, e^{-k^2 z(1-z)t} \right\} e^{-tm^2}. 
\label{eqn:iiitwo}
\end{eqnarray}

\vspace{2mm}

It was stated above that $(S \! \not\!\! A)^3$ is trace class in two
dimensions.  This follows from the result~\cite{xxvi,xxvii} that the
operator $S(P) \! \not\!\!\! A(X)$ is a bounded operator on $L^2(\text{I}\!\text{R}^2)$ in
the trace ideal ${\mathcal{C}}_n$, $n>2$ and 
\begin{equation}
\|S(P) \! \not\!\! A(X)\|_n \le \|S\|_{L^n}\|\! \not\!\! A\|_{L^n}.
\label{eqn:iiithree}
\end{equation}
Here ${\mathcal{C}}_n = \{A\,|\|A\|^n_n = \text{Tr}\,(A^\dagger
A)^{n/2}\, <
\infty\}$.
By inspection $\|S\|_{L^n} < \infty$ for $n>2$.  We hereafter assume
that $A_\mu \in L^n(\text{I}\!\text{R}^2)$, $n>2$, which is compatible with the
$1/r$ fall off of $A_\mu$ in the gauge $\partial_\mu A^\mu = 0$ when
$\Phi \not = 0$.  Since 
$S \! \not\!\! A \in {\mathcal{C}}_{2+\epsilon}$
it belongs to all ${\mathcal{C}}_n$ with $n>2$, thus establishing our
statement that 
$(S \! \not\!\! A)^3$
is trace class in two dimensions.

In the coordinate space representation of
$S(P) \! \not\!\! A(X)$
the propagator is given by

\vspace{2mm}
\begin{equation}
S(x) = \frac{1}{2\pi} (\sqrt{m^2} + i\! \! \not\! \partial ) K_0
(\sqrt{m^2x^2} ).
\label{eqn:iiifour}
\end{equation}

\vspace{2mm}
\noindent
Hence $S$ is an analytic function of $m^2$ throughout the complex
$m^2$-plane cut along the negative real axis.  Then the following
theorem of Gohberg and Kre\v{\i}n~\cite{xxviii} applies:  Let $A(\mu)
\in {\mathcal{C}}_1$ and be analytic in $\mu$ in some region.  Then
the determinant $\text{det} (1-A(\mu))$ is analytic in $\mu$ in the
same region.  In our case 
$S \! \not\!\! A \in {\mathcal{C}}_{2+\epsilon}$,
requiring the two subtractions in (\ref{eqn:iiione}).  These
subtractions can be easily incorporated into Gohberg and Kre\v{\i}n's
proof for
$S \! \not\!\! A \in {\mathcal{C}}_1$,
provided use is made of the inequality~\cite{xxiii,xxv}
\begin{equation}
|\text{det}_n(1+A)| \le e^{\Gamma_n \| A \|^n_n},
\label{eqn:iiifive}
\end{equation}
if $A \in {\mathcal{C}}_n$
and $\Gamma_n$
is a constant.  Therefore,
$\text{det}_3 (1-S \! \not\!\! A)$
is infinitely differentiable in $m^2$ on the open interval $(0,
\infty)$.  In addition, $\text{det}_3$ has no zeros for $m^2 > 0$ and
for real coupling.  This was proved in Sec. III C
of~\cite{i} for the case of $\text{det}_4$ in three dimensions; the
case of $\text{det}_3$ in two dimensions follows immediately from this
proof.  The regulated determinant, $\ln \text{det}_n$, is analogous to
(\ref{eqn:iiione}) with $n-1$ subtractions.  Hence, $\ln \text{det}_3$
is also infinitely differentiable in $m^2$ on $(0, \infty)$.

Next, we require a theorem of Ford~\cite{xxix}:  Let $f(x)$ be an
infinitely differentiable function of $x$ on $(a, \infty)$ and let
$\phi(x) = f(1/x)$.  If the limits $\phi(+0), \phi'(+0), \ldots$
exist then for $x$ on $(a, \infty)$,
\begin{equation}
f(x) \sim a_0 + a_1/x + \cdots,
\label{eqn:iiisix}
\end{equation}
with $a_k = \phi^{(k)}(+0)/k!$, $k = 0,1,\ldots$.  The series
(\ref{eqn:iiisix}) is asymptotic in the sense that

\vspace{2mm}
\begin{equation}
\lim_{x \to \infty} x^n [ f(x) - (a_0 + a_1/x + \cdots + a_n/x^n) ] =
0
\label{eqn:iiiseven}
\end{equation}

\vspace{2mm}
\noindent
for $n = 0,1, \ldots$.

Now consider the asymptotic expansion of $\ln \text{det}_3$ for large
$m^2$.  Referring to (\ref{eqn:iiitwo}), this can be obtained from the
high-temperature expansion

\vspace{2mm}
\begin{eqnarray}
\text{Tr} (e^{-[D^2 + \frac{1}{2}\sigma F]t} - e^{-P^2t})
&=& \frac{1}{4\pi t} \int d^2x \left[\,\frac{2}{3} t^2 B^2 +
\frac{2}{15}t^3B\,\partial^2B + t^4\left(\frac{1}{70}B\,\partial^4B -
\frac{2}{45}B^4\right) \right.
\nonumber \\
&\ & \left. +\, t^5\,\left(\frac{4}{63} B^2\, \partial_\mu B\, \partial^\mu B +
\frac{1}{945} B\, \partial^6B \right) + \cdots\, \right] .
\label{eqn:iiieight}
\end{eqnarray}

\vspace{2mm}
\noindent
The terms of $O(B^2)$ are an easy consequence of second-order
perturbative theory; the term $-2t^4B^4/45$ is an immediate
consequence of the Euler-Heisenberg result specialized to two
dimensions~\cite{xviii,xxx}, and the term $4t^5B^2\, \partial_\mu B\,
\partial^\mu B/63$ follows from the results in~\cite{vii,x}, again
specialized to two dimensions.  As previously noted, the second-order
terms in (\ref{eqn:iiieight}), when substituted into (\ref{eqn:iiitwo}),
will be canceled by the counterterm, giving

\vspace{2mm}
\begin{equation}
\ln \text{det}_3 = \frac{1}{90\pi m^6} \int d^2x\, B^4 -
\frac{1}{21\pi m^8} \int d^2x\, B^2\, \partial_\mu B\, \partial^\mu B
+\cdots.
\label{eqn:iiinine}
\end{equation}

\vspace{2mm}
\noindent
Note that as the expansion continues there are integrals over
increasing derivatives and powers of $B$.  The finiteness of the
coefficients of increasing powers of $1/m^2$ requires additional
conditions on $B$.  Since the remainder after summing $n$ terms in an
asymptotic series is of the order of the first neglected
term~\cite{xxix} we must impose the additional conditions $\int d^2x\,
B^4 < \infty$ and $\int d^2 x\, B^2\, \partial_\mu B\, \partial^\mu B <
\infty$ if the series in (\ref{eqn:iiinine}) is terminated at the first
term.  We have now satisfied the conditions of Ford's theorem, thereby
establishing (\ref{eqn:iiseven}) and (\ref{eqn:iieight}).

\renewcommand{\thesection}{\Roman{section}}
\renewcommand{\theequation}{\thesection.\arabic{equation}}
\section{Four Dimensional QED} \label{sec:four}
\setcounter{equation}{0}
\renewcommand{\thesection}{\arabic{section}}
\renewcommand{\theequation}{\thesection.\arabic{equation}}

In a representation in which $\gamma_5$ takes the diagonal form
$\gamma_5 = \left( \begin{array}{cc}
1 & 0 \\
0 & -1 \end{array} \right) $
the operator $- \! \not\!\! D^2$ is
\begin{equation}
D^2 + \frac{1}{2} \sigma F =
\left( \begin{array}{cc}
H_+ & 0 \\
0 & H_- \end{array} \right),
\label{eqn:ivone}
\end{equation}
where
\begin{equation}
H_\pm = (P-A)^2 - 
\text{\boldmath{$\sigma$}}
\cdot ({\bf{B}}\pm
{\bf{E}}).
\label{eqn:ivtwo}
\end{equation}
A working definition of the chiral anomaly for $\slashletter{D}$ on
noncompact manifolds is~\cite{xxxi}

\begin{eqnarray}
\nonumber
\lim_{m^2 \downarrow 0} m^2
  \text{Tr} \left[ (H_{+} + m^2)^{-1} - (H_{-} + m^2)^{-1} \right]
  &=& \frac{1}{4\pi^2} \int d^4 x \, \mathbf{E} . \mathbf{B}\\
\label{eqn:ivthree}
  &=& - \frac{1}{16\pi^2} \int d^4 x \, ^{*}F_{\mu\nu} F^{\mu\nu},
\end{eqnarray}

\noindent
with $^*F^{\mu \nu} = \frac{1}{2} \epsilon^{\mu\nu\alpha\beta}
F_{\alpha\beta}$, $\epsilon^{0123} = 1$, $F^{k0} = E^k$, and $F^{ij} =
\epsilon^{ijk}B^k$.  The $m^2 = 0$ limit in (\ref{eqn:ivthree}) gives a
generalization of the Atiyah-Singer index theorem~\cite{xxxii} to
noncompact manifolds~\cite{xxxi},

\vspace{2mm}
\begin{equation}
\frac{1}{4\pi^2} \int d^4x\, {\bf{E}}\cdot {\bf{B}}(x) = n_+ - n_- + \frac{1}{\pi} \sum_l
\mu(l) [ \delta_+^l (0) - \delta_-^l (0)],
\label{eqn:ivfour}
\end{equation}

\vspace{2mm}
\noindent
where $n_\pm$ are the number of square-integrable zero modes of
$H_\pm$; $\delta^l_\pm (0)$ are the scattering phase
shifts for $H_\pm$ as the energy tends to zero; $l$ is a
degeneracy parameter, and $\mu(l)$ is a weight factor.

Now let us specialize to the case of smooth square-integrable
background gauge fields for which all zero modes have either positive
or negative chirality.\footnote[1]{Abelian (anti) self-dual fields are
harmonic functions and so are not square-integrable on non-compact
Euclidean space-times.} Suppose they have positive
chirality. Differentiating the definiton of (\ref{eqn:iione}) of
$\text{lndet}_{\text{QED}4}$ in the representation of (\ref{eqn:ivone})
and (\ref{eqn:ivtwo}) gives

\begin{eqnarray}
m^2 \frac{\partial}{\partial m^2} \ln \text{det}_{{\text{\scriptsize{QED}}_4}}
& = &
\frac{1}{2} m^2 \text{Tr}\,[ ( H_+ + m^2)^{-1} - ( H_- + m^2)^{-1} ]
\nonumber \\
&\ & + m^2 \text{Tr}\,[ (H_- + m^2)^{-1} - (P^2 + m^2)^{-1} ] -
\frac{1}{48\pi^2} \| F\|^2,
\label{eqn:ivfive}
\end{eqnarray}

\noindent
where the second trace in (\ref{eqn:ivfive}) is defined as

\begin{equation}
\text{Tr}\,[ (H_- + m^2)^{-1} - (P^2 + m^2)^{-1}] \equiv \int^\infty_0
dt\, e^{-tm^2} \text{Tr}\,(e^{-H_- t} - e^{-P^2t} ),
\label{eqn:ivsix}
\end{equation}

\noindent
consistent with definition (\ref{eqn:iione}). From (\ref{eqn:ivthree}),
(\ref{eqn:ivfive}) reduces to

\begin{equation}
\lim_{m^2 \downarrow 0} m^2 \pdo{m^2} \text{lndet}_{\text{QED}4}
  = - \frac{1}{32\pi^2} \int d^4 x \, (^{*}F_{\mu\nu} F^{\mu\nu}
      + \frac{2}{3} F^2_{\mu\nu} )
    + \lim_{m^2 \downarrow 0} m^2 \text{Tr}
      \left[ (H_{-} + m^2)^{-1} - (P^2 + m^2)^{-1} \right].
\label{eqn:ivseven} 
\end{equation}

\noindent If 

\begin{equation}
\lim_{m^2 \downarrow 0} m^2
  \text{Tr} \left[ (H_{-} + m^2)^{-1} - (P^2 + m^2)^{-1} \right] = 0,
\label{eqn:iveight}
\end{equation}

\noindent then for $m^2 \downarrow 0$,

\begin{equation}
\text{lndet}_{\text{QED}4}
  = - \frac{1}{32\pi^2}
    \int d^4 x (^{*}F_{\mu\nu}F^{\mu\nu} + \frac{2}{3} F^2_{\mu\nu})
    \ln m^2 + R(m^2),
\label{eqn:ivnine}
\end{equation}

\noindent where $\lim_{m^2 \downarrow 0} [R(m^2)/\ln m^2] = 0$.

Is (\ref{eqn:iveight}) true?  In two dimensions with $\Phi>0$ the
square-integrable zero modes of $H_\pm = (P-A)^2 \mp B$ are confined
to the positive chirality sector~\cite{xvi}.  We then
demonstrated~\cite{xix} that $H_-$ has the property

\vspace{2mm}
\begin{equation}
\lim_{m^2 \downarrow 0} m^2 \int^\infty_0 dt\, e^{-tm^2}
  \text{Tr}\, \left( e^{-(H_{-} + B)t} - e^{-P^2t} \right) = 0.
\label{eqn:ivten}
\end{equation}

\vspace{2mm}
\noindent
Even if $\Phi$ is positive, $B$ can fluctuate in sign.  We found that
the integral in (\ref{eqn:ivten}) only developed $\ln m^2$ type
singularities as $m^2 \downarrow 0$. This and the tendency for infrared
divergences to be less severe in higher dimensions lead us to
conjecture that (\ref{eqn:iveight}) is true. In the case when the
fermion zero modes all have negative chirality the roles of $H_{+}$
and $H_{-}$ are interchanged. Thus, in either case if

\begin{equation}
\left| \int d^4 x \; ^{*}F_{\mu\nu} \; F^{\mu\nu} \right|
  > \frac{2}{3} \int d^4 x F^2_{\mu\nu},
\label{eqn:iveleven}
\end{equation}

\noindent
then (\ref{eqn:ivnine}) indicates that $\text{lndet}_{\text{QED}4}$
becomes negative as $m^2 \downarrow 0$, which is a reflection of
paramagnetism~\cite{xxii}. Since all we know is that

\begin{equation}
\int d^4 x F^2_{\mu\nu} \geq
  \left| \int d^4x \; ^{*}F_{\mu\nu} F^{\mu\nu} \right|,
\label{eqn:ivtwelve} 
\end{equation}

\noindent
(\ref{eqn:iveleven}) can not be ruled out \emph{a priori}.

Now consider the large mass behavior of
$\text{lndet}_{\text{QED}4}$. In this case
$S \slashletter{A} \in \mathcal{C}_d$, $d > 4$ provided
$A_{\mu} \in \bigcap_{n > 4} L^{n} (\mathbb{R}^4)$ so that
\cite{xii,xiii,xiv}

\vspace{2mm}
\begin{eqnarray}
\ln \text{det}_{{\text{\scriptsize{QED}}_4}} & = &
\frac{1}{8\pi^2} \int \frac{d^4k}{(2\pi)^4} |\hat{F}_{\mu\nu}(k)|^2
\int^1_0 dz\,z(1-z) \ln \left(\frac{z(1-z)k^2 + m^2}{m^2} \right) 
\nonumber \\
&\ & + \int \Pi_{\mu\nu\lambda\sigma}A_\mu A_\nu A_\lambda A_\sigma +
\ln \text{det}_5
(1 - S \! \not\!\! A).
\label{eqn:ivthirteen}
\end{eqnarray}

\vspace{2mm}
\noindent
The first two terms in (\ref{eqn:ivthirteen}) are obtained from the
definition (\ref{eqn:iione}) by expanding it through fourth order.
The fourth-order term in (\ref{eqn:ivthirteen}) has been dealt with
explicitly by Karplus and Neuman~\cite{xxxiii}.  Inspection of their
result continued to Euclidean space shows that it is analytic in the
complex $m^2$-plane cut along the negative real axis.  By Ford's
theorem~\cite{xxix} it has an asymptotic expansion in $1/m^2$, whose
leading term is~\cite{xxxiii}

\vspace{2mm}
\begin{equation}
\int \Pi_{\mu\nu\lambda\sigma}A_\mu A_\nu A_\lambda A_\sigma =
\frac{1}{2880\pi^2 m^4} \int d^4x\,[14 
F_{\mu\nu} F_{\nu\alpha} F_{\alpha\beta} F_{\beta\mu} - 5(F_{\mu\nu}
F_{\mu\nu})^2] + \cdots.
\label{eqn:ivfourteen}
\end{equation}

\vspace{2mm}

The remainder term, 
$\ln \text{det}_5$, 
in (\ref{eqn:ivthirteen}) is like 
$\ln \text{det}_3$
in (\ref{eqn:iiione}) except that it has four subtractions:
\begin{equation}
\ln \text{det}_5 (1 - S \! \not\!\! A) = \text{Tr}\, [ \ln(1 - S \! \not\!\! A)
+ \sum_{n=1}^4 (S \! \not\!\! A)^n /n ].
\label{eqn:ivfifteen}
\end{equation}

\vspace{2mm}
\noindent
In the coordinate space representation of $S \! \not\!\! A$ the
propagator is

\vspace{2mm}
\begin{equation}
S(x) = \frac{m^2}{4\pi^2} (\sqrt{m^2} + \! \not\!
\partial)(K_1(\sqrt{m^2x^2}) / \sqrt{m^2x^2} ),
\label{eqn:ivsixteen}
\end{equation}

\vspace{2mm}
\noindent
which is analytic in $m^2$ throughout the complex $m^2$-plane cut
along the negative real axis.  Therefore, the same analysis as in Sec.
\ref{sec:three} establishes that $\text{det}_5$ is infinitely
differentiable in $m^2$ on the interval $(0, \infty)$ provided use is
made of (\ref{eqn:iiifive}) for $n=5$ to extend Gohberg and Kre\v{\i}n's 
theorem to $\text{det}_5$.  Moreover, $\text{det}_5$ has no zeros for
$m^2 > 0$ for real coupling.  Again, the proof of this follows
immediately from the proof in Sec. III C of Ref.~\cite{i} that 
$\text{det}_4$ has no zeros in $\text{QED}_3$ for $m^2 > 0$ and real
coupling.  Hence, $\ln \text{det}_5$ is also infinitely differentiable
in $m^2$ on $(0, \infty)$ and will have an asymptotic expansion in
$1/m^2$ for a restricted class of fields.  By Furry's theorem and
power counting we know that the first term in its expansion will be
$O(\int d^4x \, F_{\mu\nu}^6/m^8)$. 

This leaves the first term in (\ref{eqn:ivthirteen}).  By inspection we
now have for large $m^2$

\vspace{2mm}
\begin{equation}
\ln \text{det}_{{\text{\scriptsize{QED}}_4}} = \frac{1}{240\pi^2m^2}
\int d^4x\,(\partial_\alpha F_{\mu\nu})^2 + R_2 + R_4 + R_5.
\label{eqn:ivseventeen}
\end{equation}

\vspace{2mm}
\noindent
Here $R_2$ is the remainder from the second-order term which is of
order $\int d^4x\,(\partial_\alpha \partial_\beta F_{\mu\nu})^2 / m^4$;
$R_4$ is the remainder from the fourth-order term and is of order
$\int d^4x\, F_{\mu\nu}^4/m^4$, and $R_5$, the remainder from 
$\ln \text{det}_5$, is of order $\int d^4x\,F_{\mu\nu}^6/m^8$.
Therefore, provided
$A_\mu \in \bigcap_{n > 4} L^n(\text{I}\!\text{R}^4)$
and the integrals of $(\partial_\alpha F_{\mu\nu})^2$,
$(\partial_\alpha \partial_\beta F_{\mu\nu})^2$, $F^4_{\mu\nu}$ and
$F^6_{\mu\nu}$ are finite,  $\ln \text{det}_{{\text{\scriptsize{QED}}_4}}$ 
becomes positive before dropping off to zero.  This establishes that 
$\ln \text{det}_{{\text{\scriptsize{QED}}_4}}$ 
has at least one zero for $m^2 > 0$ for the class of fields considered
if (\ref{eqn:iveight}) is true.

The existence and location of a mass zero in 
$\ln \text{det}_{{\text{\scriptsize{QED}}_4}}$ 
is renormalization dependent.  The connection between different
renormalizations is simple: if instead of subtracting at $k^2 = 0$
subtraction is made at $k^2 = \lambda^2$, (\ref{eqn:iione}) or
(\ref{eqn:ivthirteen}) give

\vspace{2mm}
\begin{eqnarray}
\ln \text{det}_{{\text{\scriptsize{QED}}_4}}(m^2, \lambda^2) = 
&&\ln \text{det}_{{\text{\scriptsize{QED}}_4}}(m^2, 0)
\nonumber \\
&\ &+ \frac{\|F\|^2}{8\pi^2} \int_0^1 dz\, z(1-z) \ln \left(
\frac{m^2}{z(1-z)\lambda^2 + m^2} \right).
\label{eqn:iveighteen}
\end{eqnarray}

\vspace{2mm}
\noindent
This trivial shift in the value of 
$\ln \text{det}_{{\text{\scriptsize{QED}}_4}}$ 
shows that a mass zero of the experimentally relevant determinant 
$\ln \text{det}_{{\text{\scriptsize{QED}}_4}}(m^2, 0)$
causes 
$\ln \text{det}_{{\text{\scriptsize{QED}}_4}}(m^2, \lambda^2)$
to reduce to a simple quadratic in the field strength.

\section{Acknowledgements} \label{sec:five}

The author wishes to thank Gerald Dunne and Chris Ford for helpful
comments.

\pagebreak


\begin{thebibliography}{99}
\bibitem{i}
M.~P.~Fry, Phys.~Rev.~D {\bf 54}, 6444 (1996).
\bibitem{ii}
M.~P.~Fry, Phys.~Rev.~D {\bf 45}, 682 (1992); D {\bf 47}, 743(E)
(1993).
\bibitem{iii}
G.~Dunne and T.~Hall, Phys.~Rev.~D {\bf 58}, 105022 (1998).
\bibitem{iv}
G.~Dunne and T.~Hall, Phys.~Rev.~D {\bf 60}, 065002 (1999).
\bibitem{v}
M.~P.~Fry, Phys.~Rev.~D {\bf 51}, 810 (1995).
\bibitem{vi}
I.~Aitchison and C.~Fraser, Phys.~Rev.~D {\bf 31}, 2605 (1985).
\bibitem{vii}
H.W.~Lee, P.Y.~Pac, and H.K.~Shin, Phys.~Rev.~D {\bf 40}, 4202 (1989).
\bibitem{viii}
D.~Cangemi, E.~D'Hoker, and G.~Dunne, Phys.~Rev.~D {\bf 51}, R2513
(1995).
\bibitem{ix}
V.P.~Gusynin and I.A.~Shovkovy, Can.~J.~Phys. {\bf 74}, 282 (1996).
\bibitem{x}
V.P.~Gusynin and I.A.~Shovkovy, J.~Math.~Phys. {\bf 40}, 5406 (1999).
\bibitem{xi}
E.~Brezin and C.~Itzykson, Phys.~Rev.~D {\bf 2}, 1191 (1970);\\
C.~Martin and D.~Vautherin, Phys.~Rev.~D {\bf 38}, 3593 (1988).
\bibitem{xii}
E.~Seiler, Phys.~Rev.~D {\bf 22}, 2412 (1980).
\bibitem{xiii}
E.~Seiler, {\it Gauge Theories: Fundamental Interactions and Rigorous
Results}, Proceedings of the International School of Theoretical
Physics, Poiana Brasov, Romania, 1981, edited by P.~Dita,
V.~Georgescu, and P.~Purice, Progress in Physics Vol.~5
(Birkh\"{a}user, Boston 1982), p.~263.
\bibitem{xiv}
E.~Seiler, {\it Gauge Theories as a Problem of Constructive Quantum
Field Theory and Statistical Mechanics}, Lecture Notes in Physics
Vol.~159 (Springer, Berlin/Heidelberg/New York, 1982).
\bibitem{xv}
J.~Schwinger, Phys.~Rev. {\bf 128}, 2425 (1962).
\bibitem{xvi}
Y.~Aharonov and A.~Casher, Phys.~Rev.~A {\bf 19}, 2461 (1979).
\bibitem{xvii}
M.~P.~Fry, Phys.~Rev.~D {\bf 53}, 980 (1996).
\bibitem{xviii}
J.~Schwinger, Phys.~Rev. {\bf 82}, 664 (1951).
\bibitem{xix}
M.~P.~Fry, J.~Math.~Phys. {\bf 41}, 1691 (2000).
\bibitem{xx}
D.~Brydges, J.~Fr\"{o}hlich, and E.~Seiler, Ann.~Phys.~(N.Y.) {\bf
121}, 227 (1979).
\bibitem{xxi}
D.~H.~Weingarten, Ann.~Phys.~(N.Y.) {\bf 126}, 154 (1980).
\bibitem{xxii}
M.~P.~Fry, Phys.~Rev.~D {\bf 55}, 968 (1997); D {\bf 56}, 6714(E),
(1997).
\bibitem{xxiii}
N.~Dunford and J.~T.~Schwartz, {\it Linear Operators}, Part II
(Interscience, New York, 1963).
\bibitem{xxiv}
E.~Seiler, Commun.~Math.~Phys. {\bf 42}, 163 (1975).
\bibitem{xxv}
B.~Simon, Adv.~Math. {\bf 24}, 244 (1977).
\bibitem{xxvi}
B.~Simon, {\it Trace Ideals and their Applications}, London
Mathematical Society Lecture Notes Series 35 (Cambridge University
Press, Cambridge, England, 1979).
\bibitem{xxvii}
E.~Seiler and B.~Simon, Commun.~Math.~Phys. {\bf 45}, 99 (1975).
\bibitem{xxviii}
I.~C.~Gohberg and M.~G.~Kre\v{\i}n, {\it Introduction to the Theory of
Linear Nonselfadjoint Operators}, Translations of Mathematical
Monographs Vol. 18 (American Mathematical Society, Providence, 1969),
p. 163.
\bibitem{xxix}
W.~B.~Ford, {\it Studies on Divergent Series and Summability and the
Asymptotic Developments of Functions Defined by Maclaurin Series}
(Chelsea, New York, 1960), p. 30.
\bibitem{xxx}
H.~Euler and B.~Kockel, Naturwissenschaften {\bf 23}, 246 (1935);
W.~Heisenberg and H.~Euler, Z.~Phys. {\bf 98}, 714 (1936);
V.~Weisskopf, K.~Dan.~Vidensk.~Selsk.~Mat.~Fys.~Medd. {\bf 14}, 6
(1936).
\bibitem{xxxi}
R.~Musto, L.~O'Raifeartaigh, and A.~Wipf, Phys.~Lett.~B {\bf 175}, 433
(1986); L.~O'Raifeartaigh, \emph{Annual Conference on Differential
Geometry in Physics}, Klausthal, August 1986 (Syracuse U. preprint SU-4228-349);
A.~Wipf, in ``\emph{Siofok 1986, Proceedings, Nonperturbative Methods in
Quantum Field Theory},'' (Balaton Conf./Workshop 1989) p.131.
\bibitem{xxxii}
M.~Atiyah and I.~Singer, Ann.~Math. {\bf 87}, 484 (1968); {\bf 87},
546 (1968); M.~Atiyah, R.~Bott, and V.~Patodi, Invent.~Math. {\bf 19},
279 (1973).
\bibitem{xxxiii}
R.~Karplus and M.~Neuman, Phys.~Rev. {\bf 80}, 380 (1950).
\end{thebibliography}
\end{document}